\documentclass[12pt,a4paper]{article}
\usepackage{geometry}
\usepackage[utf8]{inputenc}
\usepackage[T1]{fontenc}
\usepackage{amsmath,amssymb,amsthm}
\usepackage{graphicx}
\usepackage{lmodern}
\usepackage{cite}
\usepackage[linktocpage=true, colorlinks=false, linkcolor = blue, citecolor = red]{hyperref}
\begin{document}
	\title{Yet another way from field theory to gravity}
	
	\author{A. A. Sheykin\thanks{\texttt{anton.shejkin@gmail.com}}  \\
		\textit{Saint Petersburg State University, St. Petersburg, Russia}}
	\date{}
	\maketitle
	\abstract{It is shown that target space diffeomorphism invariance of a generic Lagrangian for a set of scalar fields leads to an analog of Einstein equations for the geometry of a level set of these fields.}

\section{A field theory with target space diffeomorphisms}
Consider a set of scalar fields $z^A$, $A=1...n$ in an $N$-dimensional Minkowski space with coordinates $y^a$, $a=0...N-1$, $N\geq 3$, and metric $\eta_{ab}$. Let us assume that the equations of motion for these fields are invariant w.r.t. transformation
\begin{align}\label{z'}
    z'^{A}=f^{A}(z^B),
\end{align}
where $f$ is an arbitrary function with the only restriction that the determinant of the Jacobi matrix 
\begin{align}
    J^A\,_B = \dfrac{\partial z'^A}{\partial z^B}
\end{align}
is nonzero. Such transformation could be called {\it{target space diffeomorphism}}. These transformations were studied, for example, in the low-dimensional models  in the context of integrability \cite{0712.3385} and JT gravity \cite{2202.02603}, but general properties of such transformation viewed as a gauge ones are not widely discussed in the literature. 

\section{Scalars and the form of action}
The simplest tensor w.r.t. \eqref{z'} is a derivative of $z^A$:
\begin{align}\label{dz'}
     \partial_a z'^A = J^A\,_B \partial_a z^B. 
\end{align}

This quantity allows one to construct a couple more tensors, for example:
\begin{align}\label{w}
    w^{AB} = \partial_a z^A \partial^a z^B, 
\end{align}
and its inverse $w_{AB}$ such that $w^{AC} w_{BC} = \delta^A_B$.

Obviously, it is impossible to construct a scalar w.r.t. \eqref{z'} from $w^{AB}$ alone since there is no other tensors to contract it with. However, there is another quadratic combination of derivatives of $z^A$: 
\begin{align}\label{pi}
    \pi^a_b = \partial^a z^A \partial_b z^B w_{AB}
\end{align}
which is a scalar w.r.t. \eqref{z'}. It can be easily checked that $\pi^a_c \pi^c_b=\pi^a_b$, so it has a property of a projector, and $\pi^a_{b} \delta^b_a=n$, so one cannot get rid of Lorentzian indices of $\pi$ to form a full scalar out of it.

The simplest nontrivial scalar therefore must depend on second derivatives of $z^A$. The second derivative itself is not a tensor, since
\begin{align}
    \partial_{a} \partial_{b} z'^A \equiv \partial_{ab} z'^A = J^A\,_B \partial_{ab} z^B + \partial_a z^B \partial_b z^C \dfrac{\partial^2 z'^A}{\partial z^B \partial z^C}. 
\end{align}
but it becomes one after the multiplication by 
\begin{align}
    \Pi^b_c = \delta^b_c - \pi^b_c,
\end{align}
which is also a projector, so the quantity
\begin{align}
    B^a\,_{bc} = \Pi^d_b \partial_c \Pi^a_d = - w_{AB} \partial_a z^A \Pi^d_b \partial_{cd} z^B 
\end{align}
is a scalar w.r.t. \eqref{z'}. 
It is worth noting that the role of covariant derivative is played by <<projected>> derivative:
\begin{align}
\bar{\partial}_a = \partial_a - \pi^b_a \partial_b = \Pi^b_a \partial_b.
\end{align}
All derivatives of objects which are not scalars w.r.t. \eqref{dz'} should be replaced by projected ones.

To obtain a Lorentzian scalar, one could try to contract $B_{abc}$ with itself.
There are six possible permutations of three elements, so one can expect to see six possible scalars. However, due to the easily verifiable properties \begin{align}
    \pi^a_d B_{abc} = B_{dbc}, \quad \Pi^b_d B_{abc}= B_{adc}, \quad \pi^a_c \Pi^c_b=0,
\end{align} the first and second index of $B_{abc}$ cannot be contracted, so the scalar $B_{abc} B^{bac}$ vanishes and we have the remaining five. Another possibility to construct a scalar using $B^a\,_{bc}$ is to take the derivative of it and contract the indices of the resulting object, so at the lowest nontrivial order we have seven scalars in total:
\begin{align}\label{scalars}
 \begin{gathered}
 I_1=B^a\,_{bc} B_a\,^{bc} , \  I_2=B^a\,_c\,^c B_a\,^b\,_b+B^a\,_{bc} B_a\,^{cb} \ I_3=B^a\,_c\,^c B_a\,^b\,_b-B^a\,_{bc} B_a\,^{cb}, \\ I_4=B^a\,_{bc} B^c\,_{ba}, \ I_5=B^a\,_{ba} B^c\,_{bc}, I_6=\partial_b B^{ab}\,_{a} \ I_7=\partial_b B^{ba}\,_{a}
 \end{gathered}
\end{align}
where we took a sum and a difference of two scalars for convenience.

Now we can try to write an action using these scalars. The simplest way is to take a linear combination of them:
\begin{align}\label{S1}
    S = \int d^{N} y C_i I_i,
\end{align}
where $C_i$ are constants. The variation of \eqref{S1} w.r.t. $z^A$ will give us the equations of motion. Since we are ultimately interested in physical applications of this field theory, let us impose an additional restriction: we want that these EoMs to be PDEs of no higher than second order. This  requirement is crucial, since due to the action structure, its variation has the form $b\delta b$, whereas $b = (\partial z)^2 \partial^2 z$, so after two integration by parts one could expect the presence of derivatives up to fourth order.

Let us make some comments about the calculation process. Due to the presence of five different scalars, each consisting of nonlinear combination of first and second derivatives of $z^A$, calculation by hand is deemed unreasonable. Instead, we employ \textit{Cadabra}, a computer algebra system \cite{hep-th/0701238, cs.SC/0608005,10.21105/joss.01118} which is extremely well-suited for this task \cite{1912.08839}. 
Here we will present only the results.

It turns out that the variation of the action \eqref{S1} contains third-order derivatives of $z^A$ which cannot be removed by any choice of $C_i$, so it seems like we are missing something. Let us recall that at the classical level it is not the action that should be invariant under the symmetry transformations, but rather the corresponding Euler-Lagrange equation, so we can in principle use non-scalar quantities in the action. 

The only such quantity without indices consisting of first derivatives of $z^A$ is the determinant of $w^{AB}$. From \eqref{w} and \eqref{dz'} it follows that $\det w'=|det J|^2$. Therefore we can include an arbitrary function of $w$ as a weight function in the action \eqref{S1}. Since it contains first derivatives of $z^A$, after two integrations by part such multiplier could lead to the appearance of the additional terms with derivatives of $z^A$ up to third order, which is what we want.

With that in mind, consider the following action:
\begin{align}\label{S2}
    S = \int d^{N} y f(|w|) C_i I_i.
\end{align}
This amendment makes it possible for us to obtain second-order equations.

The covariance of equations fixes the form of of the function $f(w)$ as an arbitrary power of $|w|$: otherwise different terms in the equations would transform differently w.r.t. \eqref{dz'}.

The exclusion of third- and fourth-order derivatives leads to further reduction of the action. In order to remove fourth order, one must set $C_1=C_2=0$, so the scalars $I_1=B_{abc} B^{abc}$ and $I_2=B^a\,_{bc} B_a\,^{cb}+B^a\,_c\,^c B_a\,^b\,_b$ are forbidden.

To remove the third order derivatives, one must set $C_4=-C_5$ and $C_6=-C_7$ as well as fix the power of $|w|$:
\begin{align}
    f(w)=\sqrt{|w|}.
\end{align}
After that, one could notice that the combination $I_4-I_5+I_6-I_7$ is proportional to $I_3$, so the resulting action takes the form
\begin{align}\label{action}
    S=\int d^N y \sqrt{|w|} (B^a\,_{bc} B_a\,^{cb}-B^a\,_c\,^c B_a\,^b\,_b).
\end{align}

and the corresponding equations of motion 
\begin{align}\label{EoM}
    G_{a b} B_A\,^{ab}=0,
\end{align}
where 
\begin{align}\label{Einstein}
    G_{a b} = B_{cd}\,^{d} B^{c}\,_{ab}- B_{ca}\,^d B^{c}\,_{bd} - \frac{1}{2} \Pi_{ab} (B^a\,_c\,^c B_a\,^b\,_b-B^a\,_{bc} B_a\,^{cb}),
\end{align}
and $B_A\,^{ab}=w_{AB} \partial^c z^B  B^{cab}$. 

\section{Geometric interpretation}
The action \eqref{action} and equations of motion \eqref{EoM} has a clear geometric sense. Indeed, in the scalar $I_3=B^a\,_c\,^c B_a\,^b\,_b-B^a\,_{bc} B_a\,^{cb}$ one could easily recognize the expression for scalar curvature of a surface given by Gauss equation\cite{kobno}, because $B_{abc}$ after the contraction of its third index with a tangent projector becomes a second fundamental form of certain surface. These surfaces are, in fact, the surfaces of constant values of the fields $z^A$, $w^{AB}$ is a metric of the subspace orthogonal to these surfaces, and $\Pi_{ab}$ is a tangent projector onto these surfaces, see details in \cite{statja25}. If one introduces the coordinates $x^\mu$ on these surfaces and performs the change of variables, the action \eqref{action} can be rewritten as the sum of Einstein-Hilbert actions of each surface \cite{statja42}:
\begin{align}
    S=\int d^{n} z \int d^{N-n} x \sqrt{-g} R.
\end{align}
The equations \eqref{EoM} can, in turn, be cast in the form of vacuum Regge-Teitelboim equations, which are the main equations of embedding theory approach proposed fifty years ago \cite{regge}. These equations might be treated as extension of Einsteinian dynamics \cite{statja51} and can be made equivalent to them after a suitable choice of initial data.

We therefore arrive to the following conclusion: in a lowest nontrivial order of a theory of a set of scalar fields living in Minkowski space, the invariance of the theory w.r.t. target space diffeomorphisms together with the requirement of the absence of higher order derivatives in the equations of motion turns the action of the theory to the sum of Einstein-Hilbert actions for level sets of these fields. 

There are still some questions to be answered. Firstly. in the above exposition we did not include the matter terms in the action for brevity, but it can and should be done. Secondly, in the case of two-dimensional surface in three-dimensional spacetime the cancellation of higher order derivatives could still happen if some of the constraints on the form of action is relaxed. These questions will be studied later. 

{\bf{Acknowledgements}}. The author is grateful to S. A. Paston for useful discussions.


\begin{thebibliography}{10}
\newcommand{\enquote}[1]{``#1''}
\providecommand{\url}[1]{\texttt{#1}}
\providecommand{\urlprefix}{URL }
\expandafter\ifx\csname urlstyle\endcsname\relax
  \providecommand{\doi}[1]{doi:\discretionary{}{}{}#1}\else
  \providecommand{\doi}{doi:\discretionary{}{}{}\begingroup \urlstyle{rm}\Url}\fi
\providecommand{\eprint}[1]{\href{http://arxiv.org/abs/#1}{\texttt{#1}}}

\bibitem{0712.3385}
Christoph Adam, \enquote{Integrability and Diffeomorphisms on Target Space}, \href{http://dx.doi.org/10.3842/sigma.2007.123}{\emph{SIGMA}, \textbf{3} (2007), 123}.

\bibitem{2202.02603}
Carlos Valcárcel, Dmitri Vassilevich, \enquote{Target space diffeomorphisms in Poisson sigma models and asymptotic symmetries in 2D dilaton gravities}, \href{http://dx.doi.org/10.1103/physrevd.105.106016}{\emph{Physical Review D}}, \textbf{105}: 10.

\bibitem{hep-th/0701238}
Kasper Peeters, \enquote{Introducing Cadabra: a symbolic computer algebra system for field theory problems}, 2018, \eprint{hep-th/0701238}.

\bibitem{cs.SC/0608005}
Kasper Peeters, \enquote{Cadabra: a field-theory motivated symbolic computer algebra system}, \href{http://dx.doi.org/10.1016/j.cpc.2007.01.003}{\emph{Comp. Phys. Comm.}}, \textbf{176}: 8 (2007), 550–558.

\bibitem{10.21105/joss.01118}
Kasper Peeters, \enquote{Cadabra2: computer algebra for field theory revisited}, \href{http://dx.doi.org/10.21105/joss.01118}{\emph{Journal of Open Source Software}}, \textbf{3}: 32 (2018), 1118.

\bibitem{1912.08839}
Leo Brewin, \enquote{Using Cadabra for tensor computations in General Relativity}, 2019, \eprint{1912.08839}.

\bibitem{kobno}
Sh. Kobayashi, K.~Nomizu, \enquote{Foundations of Differential Geometry}, vol. 1,2, Wiley, New York, 1963,1969.

\bibitem{statja25}
S.~A. Paston, \enquote{Gravity as a field theory in flat space-time}, \href{http://dx.doi.org/10.1007/s11232-011-0138-3}{\emph{Theor. Math. Phys.}}, \textbf{169}: 2 (2011), 1611--1619, \eprint{arXiv:1111.1104}.

\bibitem{statja42}
A.~A. Sheykin, S.~A. Paston, \enquote{Field-Theoretical Formulation of Regge-Teitelboim Gravity}, \href{http://dx.doi.org/10.1134/S1063778816100124}{\emph{Phys. At. Nucl.}}, \textbf{79}: 11 (2016), 1494, \eprint{1704.06883}.

\bibitem{regge}
T.~Regge, C.~Teitelboim, \enquote{General relativity \`a la string: a progress report}, in \emph{Proceedings of the First Marcel Grossmann Meeting, Trieste, Italy, 1975}, edited by R.~Ruffini, 77--88, North Holland, Amsterdam, 1977, \eprint{arXiv:1612.05256}.

\bibitem{statja51}
S.~A. Paston, A.~A. Sheykin, \enquote{Embedding theory as new geometrical mimetic gravity}, \href{http://dx.doi.org/10.1140/epjc/s10052-018-6474-9}{\emph{The European Physical Journal C}}, \textbf{78}: 12 (2018), 989, \eprint{arXiv:1806.10902}.

\end{thebibliography}
\end{document}